\pdfoutput=1

\documentclass[aps,prd,amsmath,floats,floatfix,twocolumn,superscriptaddress,nofootinbib,showpacs,longbibliography]{revtex4-1}

\usepackage[T1]{fontenc}
\usepackage[utf8]{inputenc}
\usepackage{lmodern}
\usepackage{mathrsfs}
\usepackage{amsthm}
\usepackage{verbatim}
\usepackage{makecell}
\usepackage{microtype}
\usepackage[dvipsnames, usenames]{xcolor}
\definecolor{linkcolor}{rgb}{0.0,0.3,0.5}
\usepackage[hypertexnames=false, unicode, colorlinks=true, linkcolor=linkcolor,
citecolor=linkcolor, filecolor=linkcolor,urlcolor=linkcolor,
pdfusetitle]{hyperref}

\usepackage[all]{hypcap}
\usepackage{graphicx}
\usepackage{xspace}
\usepackage{braket}
\usepackage{amssymb}
\usepackage[normalem]{ulem} 
\usepackage{bm} 

\usepackage{array}
\usepackage{multirow}
\usepackage{microtype}

\usepackage{lipsum}  

\usepackage{blindtext}
\usepackage{mathrsfs}
\usepackage{orcidlink}
\graphicspath{%
  {figures/}%
}

\DeclareMathAlphabet{\mathpzc}{OT1}{pzc}{m}{it}

\usepackage{mathtools}

\newcommand{\AffSCUT}{\affiliation{%
  School of Physics and Optoelectronics, South China University of Technology,
  Guangzhou 510641, China}}
\newcommand{\AffNBU}{\affiliation{%
  Institute of Fundamental Physics and Quantum Technology, and School of Physical Science
  and Technology, Ningbo University, Ningbo, Zhejiang 315211, China}}
\newcommand{\AffNBUlab}{\affiliation{%
  Zhejiang Key Laboratory of Extreme Universe, Ningbo University,
  Ningbo, Zhejiang 315211, China}}
\newcommand{\AffBNU}{\affiliation{%
  School of Physics and Astronomy, Beijing Normal University,
  Beijing 100875, China}}
\newcommand{\AffNKU}{\affiliation{%
  School of Physics, Nankai University, Tianjin 300071, China}}

\newcommand{\TITLE}{Near-extremal critical dynamics driven by a phantom field in Einstein–Maxwell–scalar theory}
\begin{document}

\title{\TITLE}


\author{Zhen-Kun Jia}
\email{202420130512@mail.scut.edu.cn}
\AffSCUT
\author{Lang Zhao}      \AffBNU
\author{Lin Chen}       \AffNBU \AffNBUlab
\author{Peng-Cheng Li}  \AffSCUT
\author{Cheng-Yong Zhang} \AffNKU
  \email{zhangcy@nankai.edu.cn}

\begin{abstract}
We investigate the fully nonlinear dynamics of Einstein--Maxwell--scalar black holes in the low-mass, near-extremal region of the solution space, which is difficult to access with conventional positive-energy perturbations. An independent phantom field is introduced as a controllable negative-energy channel, and its initial amplitude is varied while the remaining parameters of the initial data are held fixed. The phantom contribution is concentrated during the early-time infall and reduces the quasi-local mass in the black-hole region at fixed electric charge, thereby allowing the evolution to approach near-extremal configurations. Near a critical amplitude $p_\ast$, evolutions on both sides of the threshold enter the same long-lived scalarized intermediate stage before separating toward either a near-extremal Reissner--Nordstr\"om black hole or a nonextremal hairy black hole. The residence time near this intermediate stage obeys the logarithmic scaling law of Type I critical behavior, and the growth rate extracted from the departure is consistent with the inverse lifetime coefficient. The intermediate stage is characterized by a small horizon surface gravity and finite horizon curvature, with no curvature blow-up resolved at the horizon during the simulated interval. These results demonstrate that phantom-assisted negative-energy transport provides a dynamical route into the near-extremal sector of the EMS solution space and reveals a threshold-mediated transition between bald and hairy black-hole end states.

\end{abstract}

\maketitle



\section{Introduction}
\label{sec:introduction}

Black holes can acquire scalar hair when a scalar field couples nonminimally to gravity or to other matter fields~\cite{Herdeiro:2015waa,Doneva:2022ewd}. One mechanism is spontaneous scalarization: the scalar-free configuration becomes unstable, and the scalar field grows until nonlinear effects become important. Originally proposed for neutron stars~\cite{Damour:1993hw}, this mechanism has been realized for black holes through couplings to curvature and to the electromagnetic field~\cite{Doneva:2017bvd,Silva:2017uqg,Herdeiro:2018wub}. In Einstein--Maxwell--scalar (EMS) theory, the latter coupling allows charged Reissner--Nordstr\"om (RN) black holes to evolve into scalarized black holes~\cite{Fernandes:2019rez,Xiong:2022ozw}. The existence and stability of these solutions depend on the coupling function~\cite{Astefanesei:2019pfq,Myung:2019oua,Myung:2018jvi}. A different situation arises when its leading correction is quartic in the scalar field. RN black holes then remain linearly stable against scalarization, while a sufficiently strong perturbation can still trigger the formation of scalar hair~\cite{Zhang:2021nnn,Jiang:2023yyn}. The corresponding scalarized solutions have also been studied with rotation~\cite{Xiong:2023bpl}, and related nonlinear mechanisms occur in scalar--Gauss--Bonnet theories~\cite{Doneva:2021tvn,Doneva:2022yqu}. Of particular interest is the coexistence of stable RN and scalarized black holes together with an unstable scalarized branch, as found in quartic EMS models~\cite{Blazquez-Salcedo:2020nhs,LuisBlazquez-Salcedo:2020rqp,Jiang:2023yyn}. In this case, the final state depends on the nonlinear evolution of the perturbation.

Near the threshold for scalarization, an EMS black hole does not evolve directly toward its final state. Instead, it first approaches an unstable scalarized black hole, stays close to it for a long time, and eventually departs toward either an RN or a stable scalarized black hole~\cite{Zhang:2021nnn,Jiang:2023yyn}. The duration of this intermediate stage increases logarithmically as the perturbation amplitude approaches its critical value. This behavior connects black-hole scalarization to critical gravitational collapse~\cite{Choptuik:1992jv,Gundlach:2007gc}, particularly Type I criticality associated with an unstable finite-mass solution~\cite{Choptuik:1996yg,Brady:1997fj,Hawley:2000dt}. Related transitions between black-hole end states occur in Einstein--Yang--Mills theory~\cite{Choptuik:1999gh,Rinne:2014kka,Maliborski:2017jyf}. Nonlinear scalarization and descalarization have also been studied in asymptotically anti-de Sitter EMS models~\cite{Guo:2021zed,Zhang:2021etr,Luo:2022roz,Zhang:2022cmu,Chen:2023eru} and in generalized scalar--tensor gravity~\cite{Liu:2022fxy}. These studies establish the dynamical role of unstable scalarized
black holes. The structure of the static solution space also
motivates the study of slow evolution near the endpoints of the
scalarized branches. At the high-mass end, the stable and unstable
scalarized branches merge, and evolution beyond this merger can
exhibit a long-lived bottleneck associated with a saddle-node
ghost, with a power-law timescale~\cite{Zhang:2026erm}.
This observation motivates us to investigate the dynamics near
the low-mass end of the unstable branch, which approaches the
extremal RN boundary. A related limiting structure has been
studied in a quartic polynomial EMS model~\cite{Blazquez-Salcedo:2020crd}.

The question is whether evolution toward this low-mass region
also encounters a bottleneck, and what critical configuration
controls the subsequent dynamics. In particular, near the
extremal RN mass $M=Q$, it is necessary to examine whether the evolution
can approach a regular scalarized black hole or instead
develops strong curvature growth that prevents its numerical
continuation. The possibility of exposing a singularity
provides a further motivation for examining the horizon
geometry during this process.

To investigate these questions, we need a way to drive the
black hole toward the low-mass region while keeping its
electric charge fixed. Energy exchange during nonlinear hair formation can be substantial, as illustrated by EMS energy extraction and by the growth and oscillation of scalar hair in nonlinear black-hole bombs~\cite{Zhang:2025jlb,Zhang:2023qtn,Zhao:2025ijb}. For the initial-data family considered here, however, the starting point is a fixed RN black hole perturbed by a neutral scalar pulse. Increasing the amplitude of an ordinary positive-energy pulse adds energy, so it does not provide a controlled downward scan in mass~\cite{Jiang:2023yyn}. Ordinary accretion experiments probe a different direction in the EMS solution space~\cite{Corelli:2021ikv}. We therefore introduce an independent neutral phantom scalar field, whose reversed-sign kinetic term allows a negative-energy flux into the black hole. The usual horizon-growth restrictions rely on energy conditions that this field violates~\cite{Hawking:1971tu,Bardeen:1973gs,Ashtekar:2002ag}. Varying the phantom pulse amplitude thus supplies a way to lower the mass in the black-hole region without changing the electric charge. We track this local redistribution with the Misner--Sharp mass~\cite{Misner:1964je,Hayward:1994bu}, distinguishing it from the conserved total mass at spatial infinity during each evolution.

Phantom accretion has long provided a classical setting for studying black-hole mass loss~\cite{Babichev:2004yx}. Fully nonlinear simulations have followed the decrease of the black-hole mass and horizon size, and have shown that the accreted fraction depends on the initial distribution of the phantom field~\cite{Gonzalez:2009jg,Gonzalez:2016wvf}. A particularly relevant application was developed by Corelli et al.\ in Einstein--dilaton--Gauss--Bonnet gravity~\cite{Corelli:2023Evaporation,Corelli:2023Gedanken}. By adding an independent phantom scalar, they evolved negative-energy pulses that reduced the black-hole mass toward and below the minimum supported by the static black-hole solutions. Their construction makes it possible to investigate the dynamical response when mass loss drives a black hole toward the boundary of an equilibrium branch. Following this strategy, we use an independent phantom scalar
field as an auxiliary negative-energy source to access the
low-mass sector of EMS theory. Our focus is the response of
the ordinary scalar field and the resulting black-hole
dynamics near the extremal RN boundary.

In this work, we investigate the fully nonlinear evolution of
spherically symmetric, asymptotically flat EMS black holes
with an additional phantom scalar field.
Keeping the RN seed, electric charge, and ordinary-scalar
perturbation fixed, we vary the phantom pulse amplitude
to explore the dynamics near the low-mass end of the
unstable scalarized branch.
We find a threshold separating near-extremal RN and
nonextremal scalarized final states.
Near this threshold, the evolution approaches a long-lived
scalarized intermediate configuration with a small horizon
surface gravity.
The lifetime of this intermediate stage increases
logarithmically as the critical amplitude is approached,
indicating Type I critical behavior over the parameter
range studied.
This provides a comparison with the power-law bottleneck
previously found near the high-mass merger of the
scalarized branches.
We further examine the mass transport and horizon geometry
to characterize the role of the negative-energy perturbation
and the properties of the near-critical intermediate state.

The remainder of this paper is organized as follows. Section~\ref{sec:model} introduces the model, numerical formulation, initial data, and diagnostic quantities. Section~\ref{sec:results} presents the static solutions and phantom-driven evolutions, and analyzes the mass transport, critical behavior, and horizon geometry. Section~\ref{sec:conclusion} summarizes our results.

\section{Model and numerical setup}
\label{sec:model}

In units where $G=c=1$, the action is
\begin{equation}
\begin{split}S=\frac{1}{16\pi}\int d^4x\sqrt{-g}\Big[&R-2\nabla_\mu\phi\nabla^\mu\phi+2\nabla_\mu\psi\nabla^\mu\psi\\&-f(\phi)F_{\mu\nu}F^{\mu\nu}\Big],\end{split}\label{eq:action}\end{equation}
where $R$ is the Ricci scalar and $F_{\mu\nu}=\partial_\mu A_\nu-\partial_\nu A_\mu$ is the Maxwell field strength. The real scalar field $\phi$ is nonminimally coupled to the Maxwell invariant through $f(\phi)=e^{b\phi^4}$ where $b$ is the coupling parameter. The phantom field $\psi$ has no direct coupling to either $\phi$ or the Maxwell field. Its kinetic term has the opposite sign to that of $\phi$, and therefore carries negative kinetic energy.

Varying the action with respect to $g_{\mu\nu}$ gives the Einstein equations
\begin{equation}G_{\mu\nu}=2\left[T_{\mu\nu}^{\phi}+T_{\mu\nu}^{\psi}+f(\phi)T_{\mu\nu}^{A}\right],\label{eq:einstein}\end{equation}
where
\begin{align}T_{\mu\nu}^{\phi}&=\partial_\mu\phi\,\partial_\nu\phi-\frac{1}{2}g_{\mu\nu}\partial_\rho\phi\,\partial^\rho\phi,\label{eq:Tphi}\\T_{\mu\nu}^{\psi}&=-\partial_\mu\psi\,\partial_\nu\psi+\frac{1}{2}g_{\mu\nu}\partial_\rho\psi\,\partial^\rho\psi,\label{eq:Tpsi}\\T_{\mu\nu}^{A}&=F_{\mu\rho}{F_\nu}^{\rho}-\frac{1}{4}g_{\mu\nu}F_{\rho\sigma}F^{\rho\sigma}.\label{eq:TA}\end{align}
The equations of motion for the scalar and Maxwell fields are
\begin{align}\nabla_\mu\nabla^\mu\phi&=\frac{1}{4}f_{,\phi}F_{\rho\sigma}F^{\rho\sigma},\label{eq:phi_covariant}\\\nabla_\mu\nabla^\mu\psi&=0,\label{eq:psi_covariant}\\\nabla_\mu\left[f(\phi)F^{\mu\nu}\right]&=0.\label{eq:maxwell_covariant}\end{align}

\subsection{Dynamical equations}
\label{sec:dynamical_equations}

To study the fully nonlinear evolution in spherical symmetry,
we adopt the Painlev\'e--Gullstrand (PG) coordinate ansatz
\begin{equation}ds^2=-\left(1-\zeta^2\right)\alpha^2dt^2+2\zeta\alpha\,dt\,dr+dr^2+r^2d\Omega_2^2,\label{eq:pg_metric}\end{equation}
where $d\Omega_2^2$ is the line element of the unit two-sphere, and $\alpha$ and $\zeta$ are functions of $(t,r)$. This coordinate system is regular at the apparent horizon, which is located at $\zeta(t,r_h)=1$. For the RN solution,
$\alpha=1,\zeta=\sqrt{\frac{2M}{r}-\frac{Q^2}{r^2}}$,
where $M$ and $Q$ denote the mass and electric charge, respectively.
Taking the gauge potential as $A_\mu dx^\mu=A(t,r)\,dt$,
the Maxwell equation can be integrated to give
$\partial_r A=Q\alpha/[r^2f(\phi)]$.
We introduce the auxiliary variables
\begin{align}\Phi&=\partial_r\phi,&\Pi&=\frac{1}{\alpha}\partial_t\phi-\zeta\Phi,\nonumber\\\Psi&=\partial_r\psi,&P&=\frac{1}{\alpha}\partial_t\psi-\zeta\Psi.\label{eq:auxiliary_variables}\end{align}

The Einstein equations give 
\begin{align}\partial_r\zeta={}&\frac{r}{2\zeta}\left(\Phi^2+\Pi^2-\Psi^2-P^2\right)+\frac{Q^2}{2r^3\zeta f(\phi)}-\frac{\zeta}{2r}\nonumber\\&\quad+r\left(\Pi\Phi-P\Psi\right),\label{eq:zeta_constraint}\\\partial_r\alpha={}&-\frac{r\alpha}{\zeta}\left(\Pi\Phi-P\Psi\right),\label{eq:alpha_constraint}\\\partial_t\zeta={}&\frac{r\alpha}{\zeta}\Big[(\Pi+\zeta\Phi)(\Pi\zeta+\Phi)\nonumber\\
&\qquad\qquad-(P+\zeta\Psi)(P\zeta+\Psi)\Big].\label{eq:zeta_evolution}\end{align}

The equations for the scalar field can be written as
\begin{align}\partial_t\phi&=\alpha\left(\Pi+\zeta\Phi\right),\label{eq:phi_evolution}\\\partial_t\Pi&=\frac{1}{r^2}\partial_r\left[\left(\Pi\zeta+\Phi\right)\alpha r^2\right]+\frac{\alpha Q^2}{2r^4f^2(\phi)}\frac{df(\phi)}{d\phi},\label{eq:Pi_evolution}\\\partial_t\Phi&=\partial_r\left[\alpha\left(\Pi+\zeta\Phi\right)\right].\label{eq:Phi_evolution}\end{align}

The equations for the phantom field are
\begin{align}\partial_t\psi&=\alpha\left(P+\zeta\Psi\right),\label{eq:psi_evolution}\\\partial_tP&=\frac{1}{r^2}\partial_r\left[\left(P\zeta+\Psi\right)\alpha r^2\right],\label{eq:P_evolution}\\\partial_t\Psi&=\partial_r\left[\alpha\left(P+\zeta\Psi\right)\right].\label{eq:Psi_evolution}\end{align}

We fix $b=100$ and use the electric charge $Q$ as the
reference length scale, setting $Q=1$.
For the reference initial data, we take an RN seed with mass
\(M_0=1.3\), where \(M_0\) denotes the seed black-hole mass.

We initialize the scalar fields with Gaussian wave packets,
$\phi(0,r)
=
a\exp\left[-\frac{(r-16)^2}{4}\right],
\psi(0,r)
=
p\exp\left[-\frac{(r-16)^2}{4}\right],$
where \(a\) and \(p\) denote the amplitudes of the ordinary-scalar and
phantom perturbations, respectively.
We first set \(a=0\) to isolate the effect of the phantom perturbation.
For the reference near-critical family, we take \(a=0.02\) and vary
\(p\) as the control parameter.

The initial metric functions $\zeta$ and $\alpha$ are
determined by the constraint equations
\eqref{eq:zeta_constraint} and \eqref{eq:alpha_constraint}.
On subsequent time slices, $\zeta$ and the matter fields
are evolved using
Eqs.~\eqref{eq:zeta_evolution}--\eqref{eq:Psi_evolution},
while $\alpha$ is obtained from
Eq.~\eqref{eq:alpha_constraint}.
The constraint equation \eqref{eq:zeta_constraint}
is used only to construct the initial data.

The computational domain extends from an inner boundary
at $r_0=1$, located inside the horizon, to spatial
infinity. We compactify the radial coordinate as
$z=r/(r+5)$, so that spatial infinity corresponds to $z=1$.
The lapse is normalized by $\alpha\to1$ at spatial infinity.
The compactified domain is discretized uniformly using
$N=2049$ grid points. Spatial derivatives and time evolution
are evaluated using fourth-order finite differences and a
fourth-order Runge--Kutta method, respectively, with
Kreiss--Oliger dissipation employed for numerical stability.
Resolution tests confirm convergence consistent with the
fourth-order numerical scheme.

At spatial infinity, the matter fields approach their
asymptotic values and the metric function behaves as
\begin{equation}
\zeta(t,r)
=
\sqrt{\frac{2M}{r}}
\left[1+O(r^{-1})\right],
\qquad r\rightarrow\infty .
\label{eq:zeta_asymptotic}
\end{equation}
Here $M$ denotes the total mass of the system.
We evaluate it through the asymptotic limit of the
Misner--Sharp mass~\cite{Jiang:2023yyn,Corelli:2023Gedanken},
\begin{equation}
m(t,r)
=\frac{r}{2}
\left(1-g^{\mu\nu}\partial_\mu r\,\partial_\nu r\right)
=\frac{r}{2}\zeta^2(t,r),
\label{eq:misner_sharp_mass}
\end{equation}
so that
\begin{equation}
M=\lim_{r\to\infty}m(t,r)
=\lim_{r\to\infty}\frac{r}{2}\zeta^2(t,r).
\label{eq:total_mass}
\end{equation}

The apparent horizon is located at $\zeta(t,r_h)=1$.
Evaluating the Misner--Sharp mass there gives
\begin{equation}
M_h(t)
\equiv m(t,r_h(t))
=\frac{r_h(t)}{2}
=\sqrt{\frac{A_h(t)}{16\pi}},
\label{eq:horizon_mass}
\end{equation}
where $A_h(t)=4\pi r_h^2(t)$.
Here $M_h$ is the irreducible horizon mass and should
be distinguished from the total mass $M$.

\subsection{Flux and geometric diagnostics}

The time dependence of the Misner--Sharp mass provides a local
measure of radial energy transport. For a sphere of fixed areal
radius \(r\), we introduce
\begin{equation}
\mathcal{F}(t,r)
\equiv
\left.\partial_t m(t,r)\right|_r .
\label{eq:energy_flux}
\end{equation}
Thus, \(\mathcal{F}>0\) corresponds to an increase of the
quasi-local energy enclosed by the sphere, whereas
\(\mathcal{F}<0\) corresponds to a decrease.
Using Eq.~\eqref{eq:zeta_evolution}, the ordinary-scalar and
phantom contributions evaluated at the apparent horizon,
where \(\zeta_h=1\), are
\begin{align}
\mathcal{F}_{\phi,h}
&=
\alpha_h r_h^2
\left(\Pi_h+\Phi_h\right)^2
\geq 0,
\\
\mathcal{F}_{\psi,h}
&=
-\alpha_h r_h^2
\left(P_h+\Psi_h\right)^2
\leq 0.
\end{align}
The net flux is therefore
\begin{equation}
\mathcal{F}_h
=
\mathcal{F}_{\phi,h}
+
\mathcal{F}_{\psi,h}.
\end{equation}

To monitor the time-integrated flux contributions,
we define
\begin{align}
\Delta E_h^{(\phi)}(t)
&=
\int_0^t
\mathcal{F}_{\phi,h}(t')\,dt',
\\
\Delta E_h^{(\psi)}(t)
&=
\int_0^t
\mathcal{F}_{\psi,h}(t')\,dt',
\label{eq:accumulated_phantom_flux}
\end{align}
with
\begin{equation}
\Delta E_h(t)
=
\Delta E_h^{(\phi)}(t)
+
\Delta E_h^{(\psi)}(t).
\end{equation}

For the apparent horizon, we use the Hayward--Kodama
surface gravity~\cite{Hayward:1997jp},
\begin{equation}
\kappa_h
=
\left[
\frac{\partial_t\zeta}{2\alpha}
-
\zeta\,\partial_r\zeta
\right]_{r=r_h}.
\end{equation}
Using the constraint and evolution equations, this expression
reduces on the apparent horizon to
\begin{equation}
\kappa_h(t)
=
\frac{1}{2r_h(t)}
\left[
1-
\frac{Q^2 e^{-b\phi_h^4(t)}}{r_h^2(t)}
\right].
\end{equation}
For later convenience, we introduce the dimensionless horizon
quantity
\begin{equation}
\epsilon_h
\equiv
2r_h\kappa_h
=
1-
\frac{Q^2}{r_h^2 f(\phi_h)}.
\label{eq:horizon_degeneracy}
\end{equation}

We also monitor the Kretschmann scalar,
\begin{equation}K=R_{\mu\nu\rho\sigma}R^{\mu\nu\rho\sigma},\end{equation}which, in terms of the evolution variables, is given by{\small\begin{equation}\begin{aligned}K={}&8\left(\Pi^2-\Phi^2-P^2+\Psi^2\right)\left(\Pi^2-\Phi^2-P^2+\Psi^2-\frac{\zeta^2}{r^2}\right)\\&+\frac{12\zeta^4}{r^4}-16\left(\Pi\Psi-\Phi P\right)^2\\&+\frac{4Q^2e^{-b\phi^4}}{r^4}\left[\frac{5Q^2e^{-b\phi^4}}{r^4}+4\left(\Pi^2-\Phi^2-P^2+\Psi^2\right)-\frac{6\zeta^2}{r^2}\right].\end{aligned}\label{eq:kretschmann}\end{equation}}

\section{near-extremal critical dynamics}
\label{sec:results}

\subsection{Static solution space}
\label{subsec:static_solution}

\begin{figure}[t]
    \centering
    \includegraphics[width=\columnwidth]{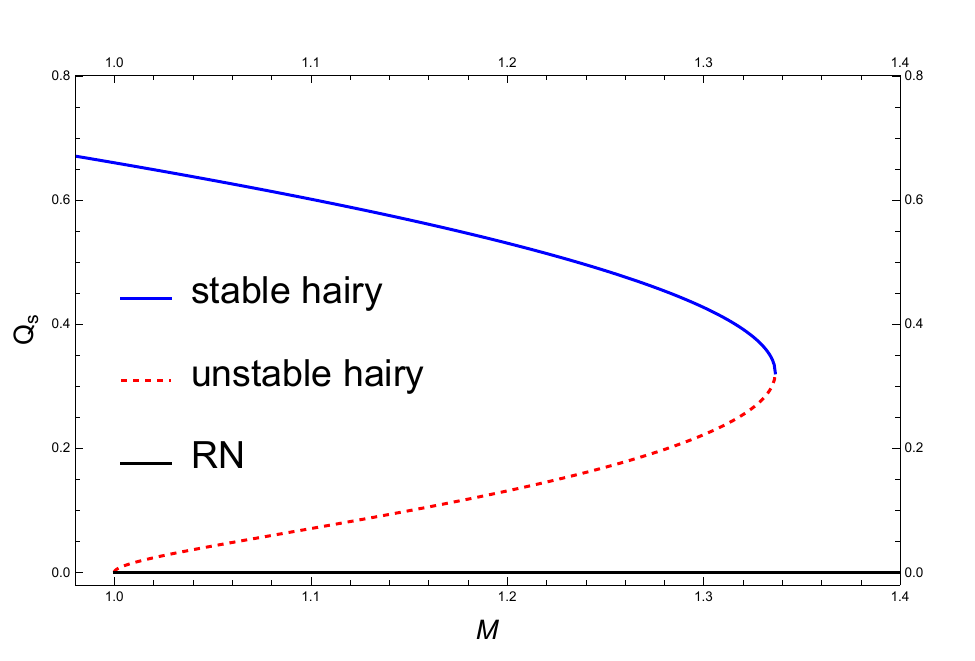}
    \caption{
    Static EMS black-hole branches for \(b=100\) and \(Q=1\)
    in the \((M,Q_s)\) plane. The black line denotes the RN branch,
    while the blue solid and red dashed curves denote the stable and
    unstable scalarized branches, respectively.
    }
    \label{fig:static_branches}
\end{figure}

Figure~\ref{fig:static_branches} shows the static EMS black-hole
solution space in the \((M,Q_s)\) plane, where the scalar charge
\(Q_s\) is defined by the asymptotic behavior
\begin{equation}
\phi(r)
=
\frac{Q_s}{r}
+
\mathcal{O}(r^{-2}).
\label{eq:scalar_charge}
\end{equation}
Besides the Reissner--Nordstr\"om branch with \(Q_s=0\), the theory
admits stable and unstable scalarized branches. The two scalarized
branches merge at the high-mass turning point, whereas the unstable
branch extends toward the extremal RN configuration on the low-mass
side.

The latter region is the focus of the present work. Since the
electric charge is kept fixed, accessing this part of the solution
space requires reducing the mass of the black-hole region. In the
following, we use the phantom field as a controllable negative-energy
source to drive the evolution toward this low-mass regime and examine
the resulting dynamics.

\subsection{Phantom-only evolution toward the low-mass regime}
\label{subsec:phantom_only}

We first consider the phantom-only case by setting \(a=0\).
For \(p=0.04\), the total mass of the initial configuration is
\(M\simeq0.94\), below the extremal RN mass for the same charge.
This evolution provides a direct test of whether negative-energy
accretion alone can drive the black hole through the low-mass
region of the EMS solution space.

Because the phantom field violates the usual energy conditions,
the apparent horizon is not constrained to grow during the
evolution and may instead contract. This raises an important issue
for the fixed inner boundary: it must remain inside a region where
all radial characteristics leave the computational domain.

For the PG metric \eqref{eq:pg_metric}, the radial null
characteristic speeds are
\begin{equation}
c_\pm(t,r)=\alpha(t,r)\bigl[-\zeta(t,r)\pm1\bigr].
\label{eq:radial_characteristic_speeds}
\end{equation}
At the inner boundary \(r=r_0\), \(c_+<0\) implies that both
radial characteristics propagate toward smaller \(r\) and leave
the computational domain. Once \(c_+>0\), an outgoing
characteristic enters through the inner boundary, and the boundary
is no longer pure outflow.

\begin{figure}[t]
    \centering
    \includegraphics[width=\columnwidth]{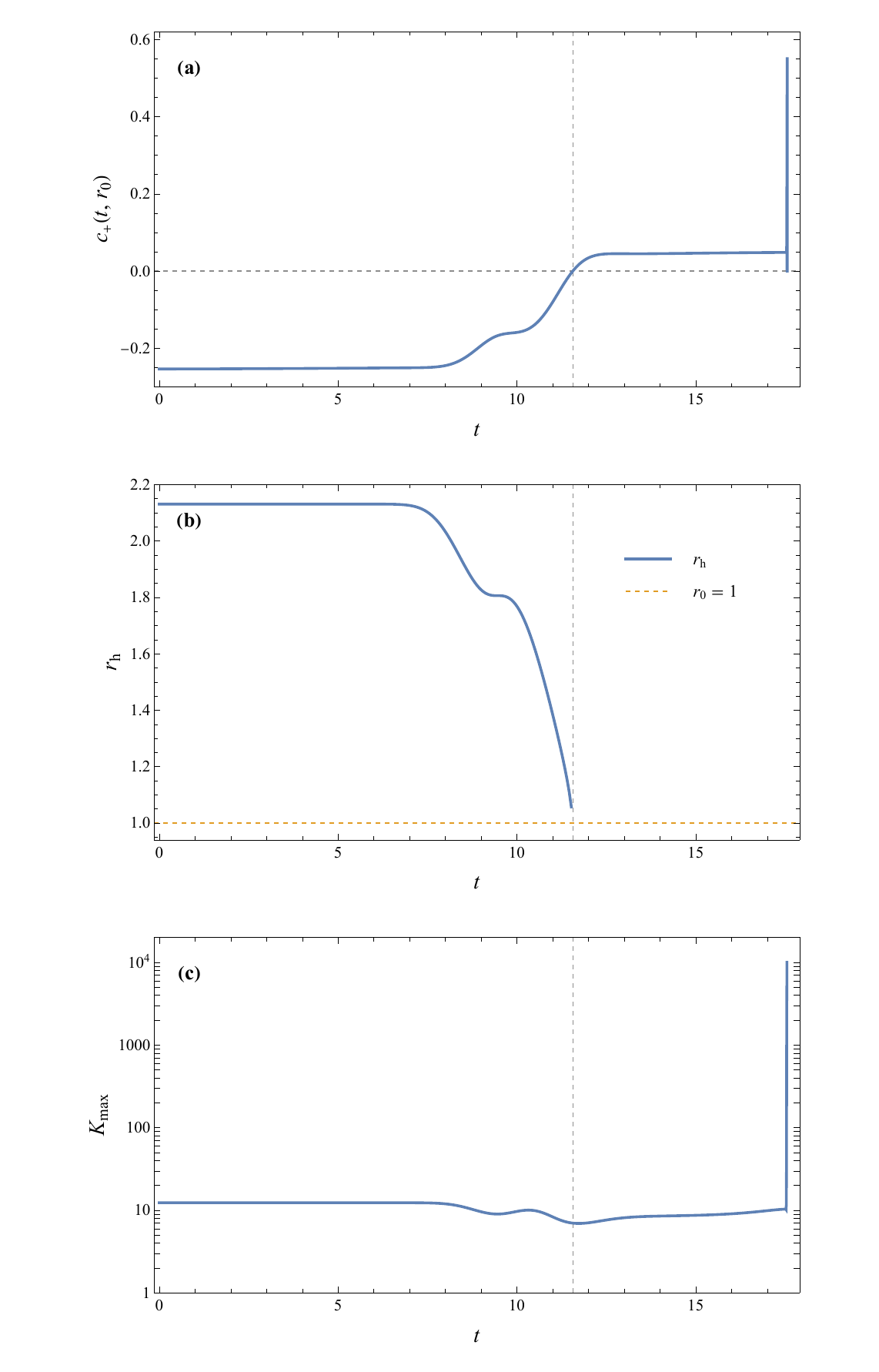}
    \caption{
Diagnostics of the phantom-only evolution with \(p=0.04\).
(a) Outgoing radial characteristic speed \(c_+(t,r_0)\) at the
fixed inner boundary. (b) Apparent-horizon radius \(r_h\), with
the horizontal dashed line marking \(r_0=1\).
(c) Maximum curvature magnitude
\(K_{\max}(t)=\max_{r\ge r_0}|K(t,r)|\).
The vertical dashed line marks
\(t_{\rm cross}\simeq11.57\), at which \(c_+(t,r_0)\) changes
sign and the inner boundary ceases to be pure outflow.
}
    \label{fig:phantom_only_curvature}
\end{figure}

Figure~\ref{fig:phantom_only_curvature} summarizes the
phantom-only evolution. The outgoing characteristic speed
\(c_+(t,r_0)\) changes sign at
\(t_{\rm cross}\simeq11.57\), while the apparent horizon
contracts toward the fixed inner boundary. At the crossing time,
the maximum curvature remains of order ten and begins to grow
rapidly only later, near \(t\simeq17.55\).

The loss of the pure-outflow condition therefore precedes the
sharp late-time growth of \(K_{\max}\). Beyond
\(t_{\rm cross}\), the present fixed-inner-boundary scheme can no
longer reliably continue the evolution without prescribing
incoming characteristic data. Consequently, the subsequent
growth of \(K_{\max}\) cannot be interpreted by itself as evidence
of a physical curvature singularity or a violation of cosmic
censorship. The physical endpoint of the phantom-only evolution
therefore remains unresolved within the present numerical setup.

Throughout the phantom-only evolution, the ordinary scalar field
remains identically zero, and no scalar hair is generated.
The phantom-only evolution therefore probes the response of the
black hole to mass reduction in the absence of scalarization, but
does not provide access to the transition between RN and
scalarized states. To investigate that transition, we next
introduce a finite ordinary-scalar perturbation and vary the
phantom amplitude \(p\) as the control parameter.

\subsection{Early-time negative-energy transport and
quasi-local mass reduction}
\label{subsec:mass_transport}

\begin{figure}[t]
    \centering
    \includegraphics[width=\columnwidth]
    {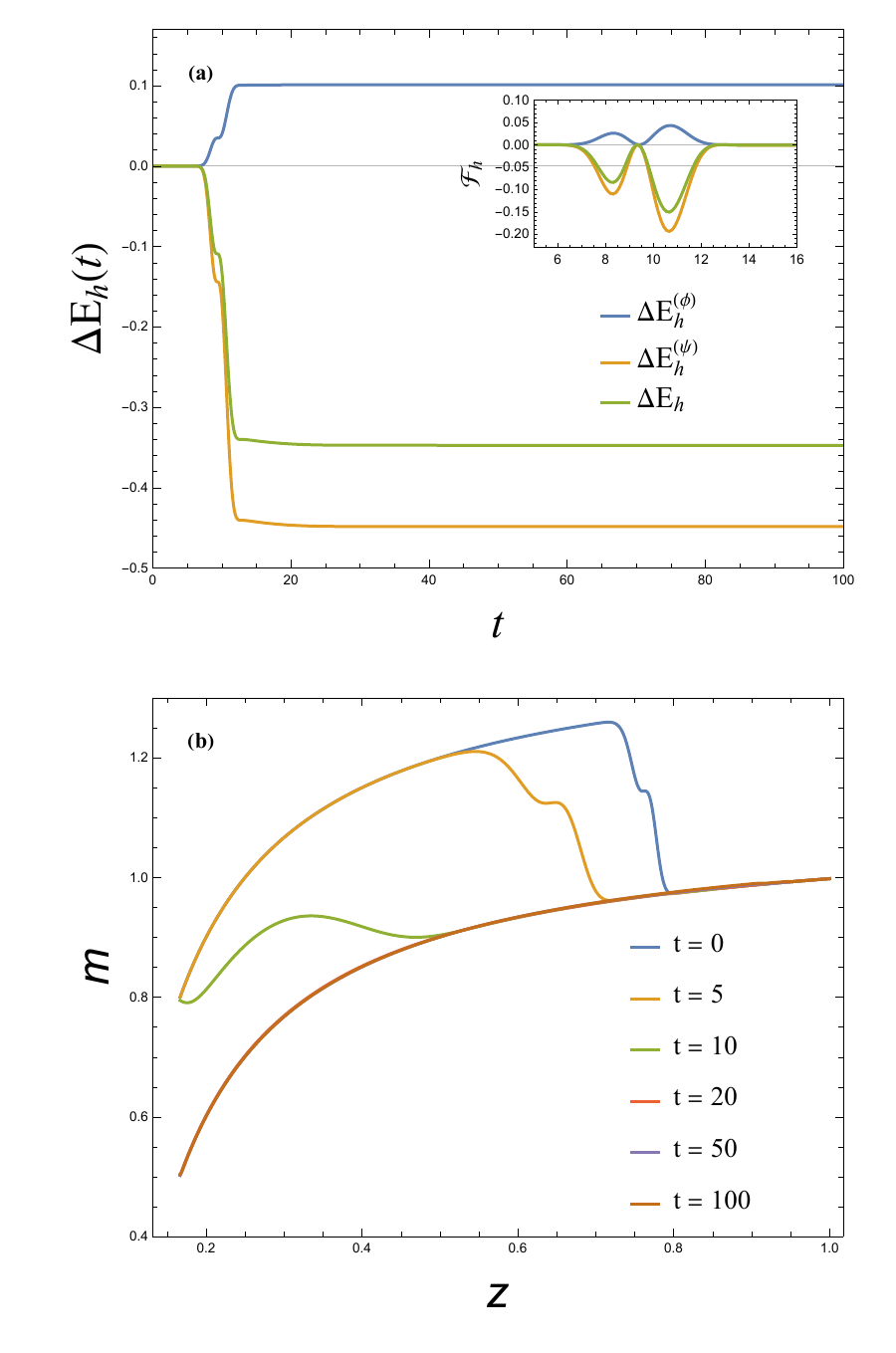}
   \caption{
Early-time energy transport and quasi-local mass redistribution
for a representative near-critical evolution with \(a=0.02\) and
\(p=0.04191778197175839\).
(a) Accumulated ordinary-scalar and phantom contributions
\(\Delta E_h^{(\phi)}\) and \(\Delta E_h^{(\psi)}\), together
with their sum \(\Delta E_h\). The inset shows the corresponding
instantaneous fluxes
\(\mathcal{F}_{\phi,h}\), \(\mathcal{F}_{\psi,h}\), and
\(\mathcal{F}_h\) at the apparent horizon.
(b) Misner--Sharp mass profiles at the times indicated in the
legend, plotted against the compactified coordinate
\(z=r/(r+5)\).
}
    \label{fig:phantom_energy_mass}
\end{figure}

We next examine the early-time mass transport for a representative
near-critical evolution of the reference family, with
\(a=0.02\) and \(p\simeq0.04191778197\).
Figure~\ref{fig:phantom_energy_mass}(a) shows the accumulated
ordinary-scalar and phantom contributions,
\(\Delta E_h^{(\phi)}\) and \(\Delta E_h^{(\psi)}\), together
with their sum \(\Delta E_h\). The inset shows the corresponding
instantaneous fluxes at the apparent horizon. During the initial
infall, the magnitude of the negative phantom contribution exceeds
the positive ordinary-scalar contribution, so that the net flux is
negative. The accumulated contributions subsequently approach
$\Delta E_h^{(\phi)}\simeq0.10,
\Delta E_h^{(\psi)}\simeq-0.45,
\Delta E_h\simeq-0.35$.

The associated redistribution of mass is shown in
Fig.~\ref{fig:phantom_energy_mass}(b). As the phantom pulse moves
inward, the Misner--Sharp mass decreases in the strong-field region.
The later profiles nearly overlap, while the asymptotic value of
\(m(t,r)\) remains unchanged. The mass reduction is therefore
quasi-local rather than a decrease of the total mass of the system.

The saturation of \(\Delta E_h^{(\psi)}\), together with the decay
of \(\mathcal{F}_{\psi,h}\), shows that the negative-energy
accretion is confined mainly to the early transient. By the time
the evolution enters the long-lived intermediate stage discussed
below, the phantom contribution at the horizon has become
negligible. The phantom field therefore acts primarily to drive the
initial quasi-local mass reduction, rather than to sustain the
subsequent near-critical evolution.

\subsection{Near-critical evolution and the long-lived
intermediate stage}
\label{subsec:near_critical}

For the reference initial-data family with \(a=0.02\), we find a
threshold in the phantom amplitude at
$p_1 \simeq 0.0419177818691396$,
which separates evolutions toward RN and scalarized black holes.
For \(p>p_1\), the evolution settles to a scalarized black hole,
whereas for \(p<p_1\), it approaches an RN black hole.

\begin{figure}[t]
    \centering
    \includegraphics[width=\columnwidth]
    {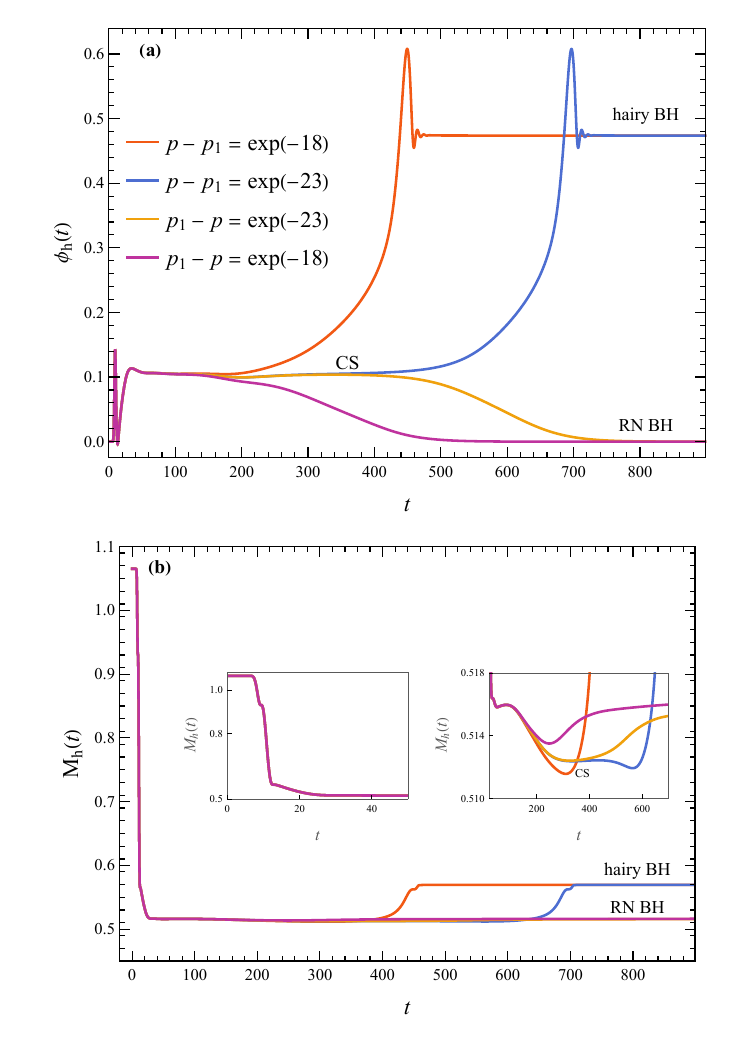}
    \caption{
Near-critical evolutions of the reference family for
\(p=p_1\pm e^{-18}\) and \(p=p_1\pm e^{-23}\).
Panel (a) shows the horizon scalar \(\phi_h(t)\), and panel (b)
shows the irreducible mass \(M_h(t)=r_h(t)/2\).
The evolutions approach a common long-lived intermediate stage
before departing toward scalarized black holes for \(p>p_1\)
or RN black holes for \(p<p_1\).
The left inset in panel (b) shows the early-time mass decrease,
while the right inset resolves the small variations of \(M_h\)
during the intermediate stage.
}
    \label{fig:near_critical_evolution}
\end{figure}

Figure~\ref{fig:near_critical_evolution} compares four near-critical
evolutions with \(p=p_1\pm e^{-18}\) and \(p=p_1\pm e^{-23}\).
After the initial transient and the quasi-local mass reduction
discussed in Sec.~\ref{subsec:mass_transport}, all four evolutions
approach a long-lived, quasistationary intermediate stage.
During this stage, the horizon scalar and irreducible mass remain
close to \(\phi_h\simeq0.10\) and \(M_h\simeq0.51\).

The evolutions subsequently depart from the intermediate stage
toward different final states. For \(p>p_1\), the horizon scalar
grows and eventually settles to a nonzero value, while \(M_h\)
increases toward that of the final scalarized black hole.
For \(p<p_1\), \(\phi_h\) decays to zero and \(M_h\) remains close
to its intermediate-stage value as the system approaches the RN
final state.
As \(p\) is tuned closer to \(p_1\), the departure from the
intermediate stage is delayed on both sides of the threshold,
while the plateau values of \(\phi_h\) and \(M_h\) remain nearly
unchanged. This behavior is consistent with the evolutions
approaching the same intermediate configuration before departing
along different directions in phase space. The dependence of the
residence time on \(|p-p_1|\) is examined in the next subsection.

\subsection{Type I critical scaling of the intermediate stage}
\label{subsec:typeI_scaling}

We now examine the lifetime of the intermediate stage and its
dependence on the distance from the critical point. To test the
robustness of the critical behavior, we consider three initial-data
families that differ in the RN seed mass \(M_0\) and the amplitude
\(a\) of the ordinary-scalar perturbation,
\begin{equation}
(M_0,a)=
\begin{cases}
(1.30,\,0.02), & p_1 \simeq 0.0419177818691396,\\
(1.18,\,0.05), & p_2 \simeq 0.0576331389779271,\\
(1.16,\,0.02), & p_3 \simeq 0.0331039627979602.
\end{cases}
\label{eq:families}
\end{equation}
All other initial-data prescriptions are the same as those specified
in Sec.~\ref{sec:model}, and the phantom amplitude \(p\) is varied
as the control parameter within each family.

Near each critical point, the evolution approaches a long-lived
intermediate configuration before eventually departing toward either
an RN or a scalarized black hole. We characterize the departure time
\(T\) by the peak of \(\ln|\dot{\phi}_h|\) associated with departure
from the intermediate stage.
As \(p\) is tuned toward \(p_*\), the departure is progressively
delayed.

\begin{figure}[t]
    \centering
    \includegraphics[width=\columnwidth]
    {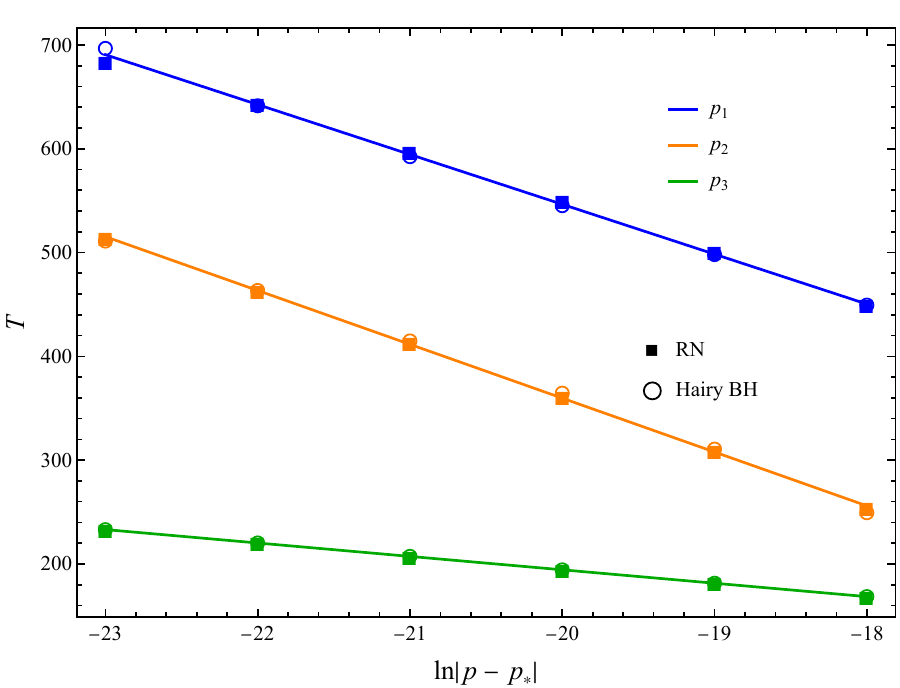}
    \caption{
Departure time \(T\) versus \(\ln|p-p_*|\) for the three
initial-data families specified in Eq.~\eqref{eq:families}.
Here \(p_*\) denotes the critical amplitude of the corresponding
family. Open circles denote evolutions toward scalarized black holes,
while filled squares denote evolutions toward RN black holes.
}
\label{fig:lifetime_scaling}
    \label{fig:typeI_scaling}
\end{figure}

Figure~\ref{fig:lifetime_scaling} shows that the lifetime of the
intermediate state follows the logarithmic scaling
$T \propto -\gamma \ln|p-p_*|$,
characteristic of Type I critical behavior.
For the reference family, the linear fit gives
$\gamma \simeq 48.74$. The logarithmic scaling follows naturally if departure from the
intermediate state is governed by a single unstable mode. For the
reference family, the leading unstable perturbation can be written as
\begin{equation}
\delta X(t)
\simeq
C\,(p-p_1)e^{\eta t},
\label{eq:unstable_mode}
\end{equation}
where \(\eta>0\) is the growth rate. The evolution leaves the
intermediate state when the unstable perturbation reaches a fixed
amplitude, which gives
$T=-\eta^{-1}\ln|p-p_1|+\mathrm{const}$.
This relation predicts $\gamma=\eta^{-1}$.
A Prony analysis of the departure of the horizon scalar yields
$\eta=0.0208\pm0.0002$,
while the lifetime fit gives
$\gamma^{-1}\simeq0.02052$.
The two estimates differ by approximately \(1.4\%\), providing an
independent consistency check that departure from the intermediate
state is governed by an unstable mode.

The two additional families exhibit the same logarithmic scaling
near their respective critical points. The Type I behavior is
therefore not restricted to the reference initial data. We emphasize,
however, that the common scaling form does not by itself imply that
all three families approach an identical intermediate configuration;
the corresponding lifetime coefficient is determined by the unstable
mode of the state approached by each family.

\subsection{Geometric characterization of the intermediate state}
\label{subsec:geometry}

\begin{figure}[t]
\centering
\includegraphics[width=\columnwidth]{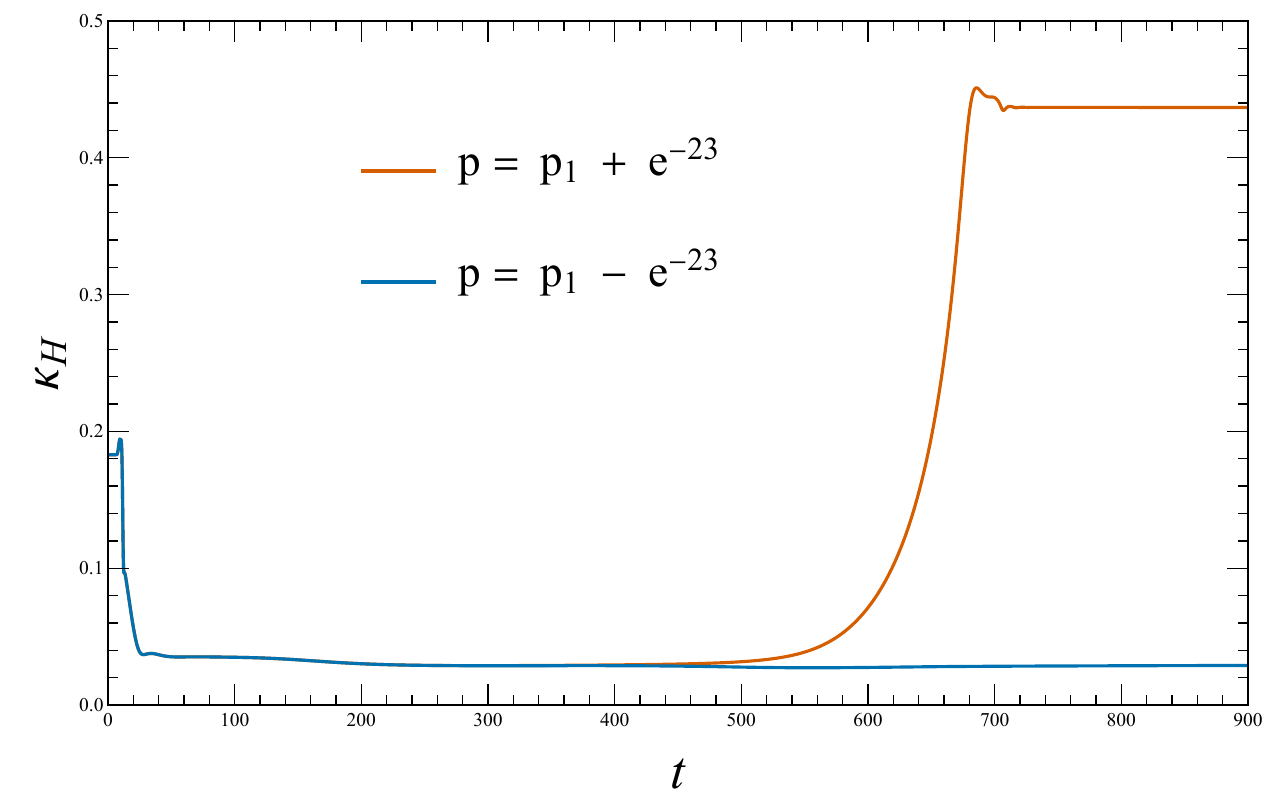}
\caption{
Hayward--Kodama surface gravity at the apparent horizon
for the near-critical evolutions of the reference family with
$p=p_1+e^{-23}$ and $p=p_1-e^{-23}$.
After the initial transient, both evolutions approach a common
quasistationary plateau with $\kappa_h\simeq0.03$.
}
\label{fig:surface_gravity}
\end{figure}

We next examine the geometry of the intermediate state for the
reference family. Figure~\ref{fig:surface_gravity} shows the
Hayward--Kodama surface gravity at the apparent horizon for
\(p=p_1\pm e^{-23}\). After the initial transient, both evolutions
approach a common plateau with $\kappa_h\simeq0.03$.
Together with the plateau value \(r_h=2M_h\simeq1.02\), this gives
$\epsilon_h=2r_h\kappa_h\simeq0.06$.
The small but positive value of \(\epsilon_h\) shows that the
intermediate state possesses a near-degenerate, but nonextremal,
apparent horizon.
Equation~\eqref{eq:horizon_degeneracy} shows that the small plateau
value of \(\epsilon_h\) measures the proximity of
\(r_h^2 f(\phi_h)\) to \(Q^2\). In the RN limit, where
\(f(0)=1\), the degenerate-horizon condition reduces to
\(r_h=|Q|\). For the scalarized intermediate state, the factor
\(f(\phi_h)\) accounts explicitly for the scalar hair.

The origin of this near-extremal geometry can be understood from
the early-time mass transport discussed in
Sec.~\ref{subsec:mass_transport}. The phantom pulse lowers the
quasi-local mass in the strong-field region while leaving the
electric charge unchanged, thereby driving the system toward the
low-mass part of the static solution space where the unstable
scalarized branch approaches the extremal RN limit.

The phantom field does not sustain this near-extremal plateau.
As shown in Sec.~\ref{subsec:mass_transport}, its horizon flux has
already become negligible by the time the intermediate state is
reached. Fine-tuning \(p\) toward \(p_1\) instead reduces the
initial amplitude of the unstable mode and increases the residence
time near the same intermediate state. The critical limit
\(p\rightarrow p_1\) therefore need not coincide with the
geometric extremal limit \(\kappa_h\rightarrow0\).

\begin{figure}[t]
    \centering
    \includegraphics[width=\columnwidth]
    {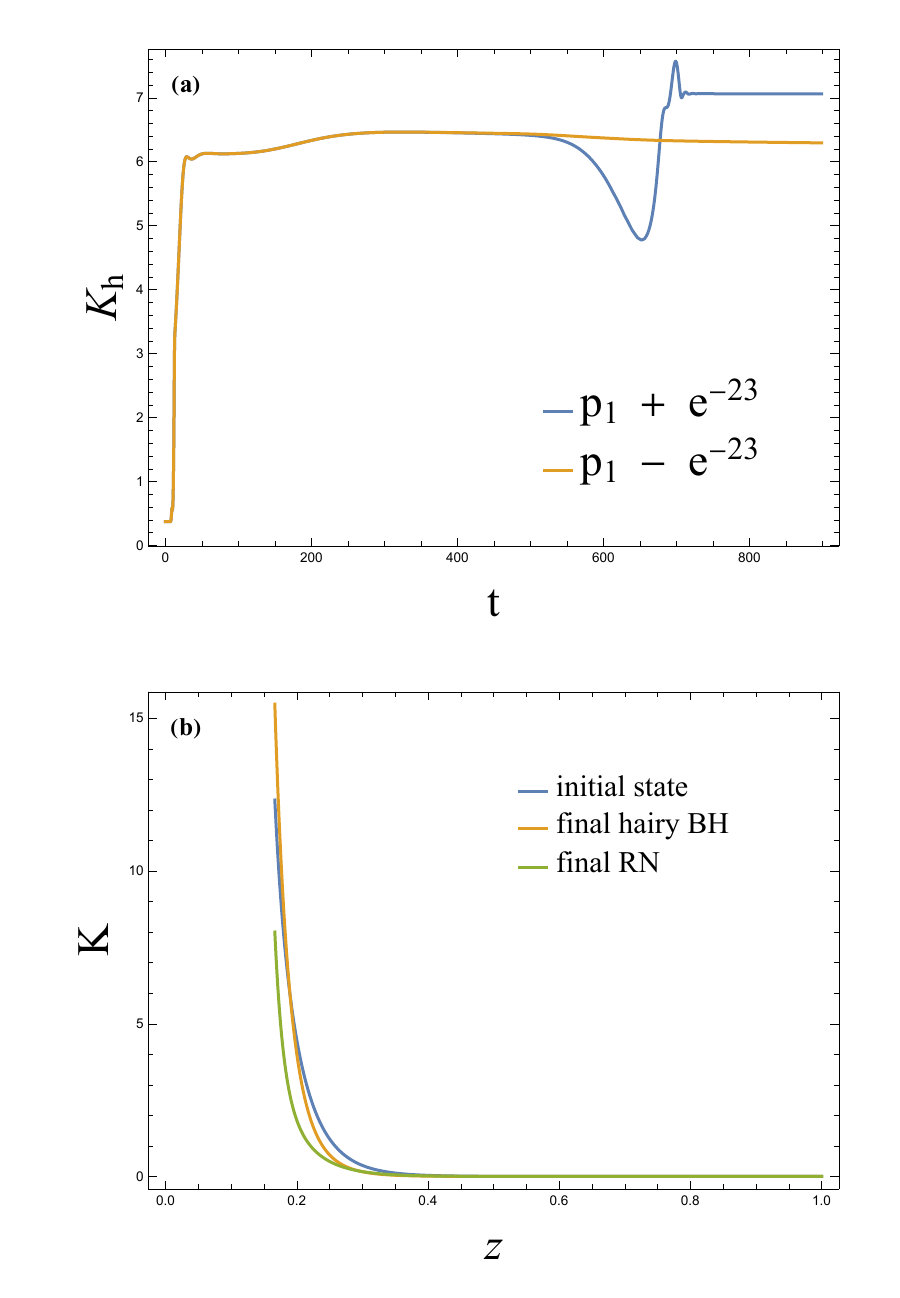}
   \caption{
Kretschmann scalar for the near-critical evolutions of the
reference family.
(a) Horizon curvature
\(K_h(t)=K(t,r_h(t))\) for \(p=p_1\pm e^{-23}\).
Both evolutions approach a common plateau with
\(K_h\simeq6.4\).
(b) Radial curvature profiles for the initial configuration and
the final scalarized and RN black holes.
}
    \label{fig:kretschmann}
\end{figure}

The horizon curvature provides an independent diagnostic.
Figure~\ref{fig:kretschmann}(a) shows that both
near-critical evolutions approach a common plateau with
$K_h \simeq 6.4$,
with no resolved secular growth during the intermediate stage.
For \(p>p_1\), departure from the plateau is followed by an
overshoot and relaxation toward the scalarized final state,
whereas for \(p<p_1\), \(K_h\) changes only weakly as the system
approaches the RN final state. The radial profiles in
Fig.~\ref{fig:kretschmann}(b) remain finite and decay
toward zero at large radius. Together with the positive surface
gravity, these results indicate that the intermediate state has a
regular, nonextremal horizon with finite curvature over the
simulated interval.

The logarithmic residence time and nonzero unstable-mode growth
rate support a Type~I interpretation of the long-lived intermediate
stage. These diagnostics distinguish the observed slowing down
from the saddle-node-induced power-law bottleneck previously
reported at the high-mass merger of the scalarized
branches~\cite{Zhang:2026erm}.

\section{Conclusion}
\label{sec:conclusion}

We have investigated the nonlinear evolution of charged
black holes near the low-mass end of the unstable
scalarized branch in Einstein--Maxwell--scalar theory,
with an additional phantom scalar field.
For the initial data considered here, a phantom
perturbation alone does not generate scalar hair:
the ordinary scalar remains identically zero when
its initial field and momentum vanish.
When both fields are perturbed, the negative-energy
pulse reduces the quasi-local mass in the strong-field
region at fixed electric charge, allowing the evolution
to probe the near-extremal regime.
Its contribution to the local horizon mass transport
becomes negligible before the long-lived intermediate
stage.

Varying the phantom amplitude while keeping the ordinary
scalar perturbation fixed reveals a threshold between
RN and scalarized final states.
For the reference family, near-threshold evolutions
approach a common, long-lived scalarized intermediate
configuration before settling into a near-extremal RN
black hole for $p<p_1$ or a nonextremal scalarized
black hole for $p>p_1$.
Over the tuning range studied, the residence time follows
$T=-\gamma\ln|p-p_1|+\mathrm{const}$,
with $\gamma\simeq48.74$.
The growth rate extracted from the departure agrees
with $\gamma^{-1}$ to approximately $1.4\%$,
supporting the interpretation of the intermediate
configuration as an unstable critical solution.
Logarithmic lifetime scaling is also observed for
two additional initial-data families.

The geometric diagnostics identify a near-degenerate
but nonextremal apparent horizon during the intermediate
stage, with $\epsilon_h\simeq0.06$ and no resolved
secular growth of the horizon Kretschmann scalar.
These results provide numerical evidence for Type I
critical behavior governing the transition between
RN and scalarized end states in the low-mass regime
studied here.

The absence of a saddle-node-induced bottleneck in these evolutions
can be understood by distinguishing the low-mass limiting structure
from the high-mass fold. Although the unstable scalarized branch
approaches the RN branch in the $(M,Q_s)$ diagram, an apparent
meeting in this projection does not by itself establish a
saddle-node bifurcation. At the high-mass turning point, the stable
and unstable scalarized equilibria merge and disappear, leaving a
saddle-node ghost that delays the subsequent
evolution~\cite{Zhang:2026erm}. Here, the observed dynamics instead
supports an unstable scalarized intermediate state from which the
evolution departs with a finite growth rate. It does not show the
equilibrium-annihilation mechanism responsible for the high-mass
bottleneck.

Fine-tuning $p$ toward $p_1$ reduces the initial unstable-mode
amplitude and thus prolongs the time required to reach a fixed
departure amplitude, yielding
$T=-\eta^{-1}\ln|p-p_1|+\mathrm{const}$.
The long-lived plateau therefore reflects Type~I critical slowing
down; no saddle-node-induced power-law bottleneck is observed
over the parameter range investigated. This conclusion applies
to the near-extremal regime studied here and does not determine
the bifurcation structure or dynamics at the exactly extremal
endpoint.

\section*{Acknowledgments}
This work is supported by the National Natural Science Foundation of China (NSFC).
L.C. acknowledges support from Grant No.~12305080, C.-Y.Z. from Grant No.~12375048,
and P.-C.L. from Grant No.~12205104.
\bibliography{References}

\end{document}